\documentclass[12pt,epsf,epsfig,wrapfig]{article}
\usepackage{amsmath,amssymb}
\allowdisplaybreaks
\begin{document}

\begin{center}
{\bfseries INTERACTING RARITA--SCHWINGER FIELD AND ITS SPIN-PARITY
CONTENT}

\vskip 5mm \underline{A.E. Kaloshin}$^{1 \dag}$ and V.P.
Lomov$^{1}$

\vskip 5mm {\small (1) {\it Irkutsk State University}\\
$\dag$ {\it E-mail: kaloshin@physdep.isu.ru }}
\end{center}

\vskip 5mm
\begin{abstract}
We obtain in analytical form the dressed propagator of the massive
Rarita-Schwinger field and discuss its properties.  The
calculation of the self-energy contributions demonstrates that
besides $s=3/2$ component the Rarita-Schwinger field contains also
two $s=1/2$ components of opposite parity.
\end{abstract}

\vskip 8mm

\noindent {\bf 1.}\ \  The vector-spinor Rarita-Schwinger field
$\Psi^{\mu}$ \cite{Rar-Sch} is used for description of the
spin-3/2 particles in QFT. However, in addition to spin-3/2 this
field contains extra spin-1/2 components and it generates the main
difficulties in its description \cite{JohnSud61,VeZw69}.

There are 10 components in decomposition of propagator so the
construction of a dressed propagator is a rather complicated issue
and its total expression is unknown up to now. Thus a practical
use of $G^{\mu\nu}$ (e.g. in case of $\Delta(1232)$ production)
needs some approximations in its description. The standard
approximation \cite{Pas95,KS} consist in a dressing the spin-3/2
components only while the rest ones can be neglected or considered
as bare. Another way to take into account the spin-1/2 components
is a numerical solution of the appearing system of
equations\cite{Kor,AKS}.

Here we derive an analytical expression for the interacting R.--S.
field's propagator with accounting all spin components and discuss
its properties. It turned out that the spin-1/2 part of the
dressed propagator has rather compact form, and a crucial point
for its deriving  is the choosing of a
suitable basis \cite{KL}.\\[0mm]

\noindent {\bf 2.}\ \ The Dyson-Schwinger equation for the
propagator of the R.--S. field has the following form
\begin{equation}
G^{\mu\nu}=G^{\mu\nu}_{0}+G^{\mu\alpha}J^{\alpha\beta}G^{\beta\nu}_{0}.
\end{equation}
Here $G_{0}^{\mu\nu}$ and $G^{\mu\nu}$ are the free and full
propagators respectively, $J^{\mu\nu}$ is a self-energy
contribution. The equation may be rewritten for inverse
propagators as
\begin{equation}
(G^{-1})^{\mu\nu}=(G^{-1}_{0})^{\mu\nu}-J^{\mu\nu}.\label{inverD}
\end{equation}
If we consider the self-energy $J^{\mu\nu}$ as a known value (so
called "rainbow" approximation), than the problem is reduced to
reversing of relation \eqref{inverD}.

The most convenient basis for the spin-tensor $S^{\mu\nu}(p)$ is
constructed by combining 5 well known  tensor operators
\cite{PvanNie,BDM,Pas95}
\begin{flalign}
(\mathcal{P}^{3/2})^{\mu\nu}=&g^{\mu\nu}-\frac{2}{3}\frac{p^{\mu}p^{\nu}}{p^2}-
     \frac{1}{3}\gamma^{\mu}\gamma^{\nu}+\frac{1}{3p^2}(\gamma^{\mu}p^{\nu}-
                                     \gamma^{\nu}p^{\mu})\hat{p},       \notag\\
(\mathcal{P}^{1/2}_{11})^{\mu\nu}=&\frac{1}{3}\gamma^{\mu}\gamma^{\nu}-
     \frac{1}{3}\frac{p^{\mu}p^{\nu}}{p^2}-\frac{1}{3p^2}(\gamma^{\mu}p^{\nu}-
     \gamma^{\nu}p^{\mu})\hat{p}, \ \ \ \ \ \
(\mathcal{P}^{1/2}_{22})^{\mu\nu}=\frac{p^{\mu}p^{\nu}}{p^2},    \notag\\
(\mathcal{P}^{1/2}_{21})^{\mu\nu}=&\sqrt{\frac{3}{p^2}}
     \cdot\frac{1}{3p^2}(-p^{\mu}+\gamma^{\mu}\hat{p}) \hat{p}p^{\nu},\ \ \ \
(\mathcal{P}^{1/2}_{12})^{\mu\nu}=\sqrt{\frac{3}{p^2}}\cdot\frac{1}{3p^2}p^{\mu}(-p^{\nu}+
\gamma^{\nu}\hat{p}) \hat{p}  \label{oper}
\end{flalign}
and off-shell projection operators
$\Lambda^{\pm}=(1\pm\hat{p}/\sqrt{p^2})/2$. Ten elements of this
basis look as
      \begin{flalign}
      \mathcal{P}_{1}=&\Lambda^{+}\mathcal{P}^{3/2},\,&
      \mathcal{P}_{3}=&\Lambda^{+}\mathcal{P}^{1/2}_{11},\,&
      \mathcal{P}_{5}=&\Lambda^{+}\mathcal{P}^{1/2}_{22},\,&
      \mathcal{P}_{7}=&\Lambda^{+}\mathcal{P}^{1/2}_{21},\,&
      \mathcal{P}_{9}=&\Lambda^{+}\mathcal{P}^{1/2}_{12},\notag\\
      \mathcal{P}_{2}=&\Lambda^{-}\mathcal{P}^{3/2},\,&
      \mathcal{P}_{4}=&\Lambda^{-}\mathcal{P}^{1/2}_{11},\,&
      \mathcal{P}_{6}=&\Lambda^{-}\mathcal{P}^{1/2}_{22},\,&
      \mathcal{P}_{8}=&\Lambda^{-}\mathcal{P}^{1/2}_{21},\,&
      \mathcal{P}_{10}=&\Lambda^{-}\mathcal{P}^{1/2}_{12},
      \label{L-basis}
      \end{flalign}
where tensor indices are omitted. We will call (\ref{L-basis}) as
the $\Lambda$-basis.

 Decomposition of a spin-tensor in this basis has the
following form:
      \begin{equation}
      S^{\mu\nu}(p)=\sum_{i=1}^{10}\mathcal{P}^{\mu\nu}_{i}\bar{S}_{i}(p^2).
      \label{l-expan}
      \end{equation}
The $\Lambda$-basis has very simple multiplicative properties
which are represented in the Table~\ref{tt}.
      \begin{table}[ht]
      \begin{tabular}{p{4mm}|p{4mm}p{4mm}p{4mm}p{4mm}p{4mm}p{4mm}p{4mm}p{4mm}p{4mm}p{4mm}}
      \qquad           & $\mathcal{P}_1$ & $\mathcal{P}_2$ &
      $\mathcal{P}_3$ & $\mathcal{P}_4$ & $\mathcal{P}_5$ &
      $\mathcal{P}_6$ & $\mathcal{P}_7$ & $\mathcal{P}_8$ &
      $\mathcal{P}_9$ & $\mathcal{P}_{10}$\\
      \hline
      $\mathcal{P}_1$    & $\mathcal{P}_1$ & 0&0&0&0&0&0&0&0&0\\
      $\mathcal{P}_2$    &0&$\mathcal{P}_2$&0&0&0&0&0&0&0&0\\
      $\mathcal{P}_3$    &0&0&$\mathcal{P}_3$&0&0&0&$\mathcal{P}_7$&0&0&0\\
      $\mathcal{P}_4$    &0&0&0&$\mathcal{P}_4$&0&0&0&$\mathcal{P}_8$&0&0\\
      $\mathcal{P}_5$    &0&0&0&0&$\mathcal{P}_5$&0&0&0&$\mathcal{P}_9$&0\\
      $\mathcal{P}_6$    &0&0&0&0&0&$\mathcal{P}_6$&0&0&0&$\mathcal{P}_{10}$\\
      $\mathcal{P}_7$    &0&0&0&0&0&$\mathcal{P}_7$&0&0&0&$\mathcal{P}_3$\\
      $\mathcal{P}_8$    &0&0&0&0&$\mathcal{P}_8$&0&0&0&$\mathcal{P}_4$&0\\
      $\mathcal{P}_9$    &0&0&0&$\mathcal{P}_9$&0&0&0&$\mathcal{P}_5$&0&0\\
      $\mathcal{P}_{10}$ &0&0&$\mathcal{P}_{10}$&0&0&0&$\mathcal{P}_6$&0&0&0\\
      \end{tabular}
\hspace{1.5cm}
\begin{tabular}{p{4mm}|p{4mm}p{4mm}p{4mm}p{4mm}}
{}&$\mathcal{P}_{1}$&$\mathcal{P}_{2}$&$\mathcal{P}_{3}$&$\mathcal{P}_{4}$\\
\hline
$\mathcal{P}_{1}$&$\mathcal{P}_{1}$&0&$\mathcal{P}_{3}$&0\\
$\mathcal{P}_{2}$&0&$\mathcal{P}_{2}$&0&$\mathcal{P}_{4}$\\
$\mathcal{P}_{3}$&0&$\mathcal{P}_{3}$&0&$\mathcal{P}_{1}$\\
$\mathcal{P}_{4}$&$\mathcal{P}_{4}$&0&$\mathcal{P}_{2}$&0\\
\end{tabular}
\\
\parbox[t]{0.55\textwidth}{\caption{Multiplicative properties of the
$\Lambda$-basis}\label{tt}} \hfill
\parbox[t]{0.4\textwidth}{\caption{Properties of basis (\ref{p4-expan})}\label{mult1}}
      \end{table}

Let us denote the inverse dressed and free propagators by
$S^{\mu\nu}$ and $S^{\mu\nu}_{0}$ respectively. Decomposing the
$S^{\mu\nu}$, $S_{0}^{\mu\nu}$ and $J^{\mu\nu}$ in $\Lambda$-basis
according to \eqref{l-expan} we reduce the equation \eqref{inverD}
to set of equations for the scalar coefficients
\begin{equation*}
\bar{S}_{i}(p^2)=\bar{S}_{0i}(p^2)-\bar{J}_{i}(p^2), \ \ \ \ \
i=1\dots 10 .
\end{equation*}

After it the reversing of the $S^{\mu\nu}$ leads to equations for
the coefficients $\bar{G}_{i}$:
\begin{equation}
\Big(\sum_{i=1}^{10}\mathcal{P}^{\mu\nu}_{i}\cdot\bar{G}_{i}(p^2)\Big)\cdot
\Big(\sum_{k=1}^{10}\mathcal{P}^{\mu\nu}_{k}\cdot\bar{S}_{k}(p^2)\Big)=
\sum_{i=1}^{6}\mathcal{P}^{\mu\nu}_{i} ,
 \label{ddecomp}
\end{equation}
which are easy to solve due to simple multiplicative properties of
$\mathcal{P}^{\mu\nu}_{i}$:
\begin{flalign}
\bar{G}_1=1/\bar{S}_1,\quad \bar{G}_3=\bar{S}_6/\Delta_1,\quad
\bar{G}_5=\bar{S}_4/\Delta_2,\quad
\bar{G}_7=-\bar{S}_7/\Delta_1,\quad
\bar{G}_9=-\bar{S}_9/\Delta_2, \notag\\
\bar{G}_2=1/\bar{S}_2, \quad \bar{G}_4=\bar{S}_5/\Delta_2,\quad
\bar{G}_6=\bar{S}_3/\Delta_1, \bar{G}_8=-\bar{S}_8/\Delta_2,\quad
\bar{G}_{10}=-\bar{S}_{10}/\Delta_1, \label{solve}
\end{flalign}
where  $\Delta_1=\bar{S}_{3}\bar{S}_{6}-\bar{S}_{7}\bar{S}_{10}$,\
\  $\Delta_2=\bar{S}_{4}\bar{S}_{5}-\bar{S}_{8}\bar{S}_{9}$.

The $\bar{G}_{1}$, $\bar{G}_{2}$ terms which describe the spin-3/2
have the usual resonance form, the $\bar{G}_{3} - \bar{G}_{10}$
terms correspond to the spin-1/2 contributions.\\

\noindent {\bf 3.}\ \
 The obtained dressed propagator of the R.--S. field has
rather unusual structure, so we would like to clarify its physical
meaning. We suggest to consider the dressing of Dirac fermions
with aim to find some analogy for R.--S. field case. The use of
the projection operators $\Lambda^\pm$ is very convenient here.

\noindent {\bf 3.1}\ \ The dressed fermion  propagator $G(p)$ is
solution of the Dyson-Schwinger equation
\begin{equation}
G(p)=G_{0}+G\Sigma G_{0}, \label{one-f-DS}
\end{equation}
where $G_{0}$  is the bare propagator and $\Sigma$  is the
self-energy contribution.

Decomposition of any matrix $4\times 4$, depending on one momentum
$p$, has the form:
\begin{equation}
S(p)=\sum_{M=1}^{2}\mathcal{P}_{M}\bar{S}^{M},\ \ \ \
\mathcal{P}_{1}=\Lambda^{+}, \ \  \qquad
\mathcal{P}_{2}=\Lambda^{-}. \label{p2-expan}
\end{equation}

Dyson-Schwinger equation in this basis takes the form:
\begin{equation}
\bar{G}^{M}=\bar{G}^{M}_{0}+ \bar{G}^{M} \bar{\Sigma}^{M}
\bar{G}^{M}_{0},   \qquad  M=1,2 . \label{tfDS-comp}
\end{equation}

Let us look at the self-energy contribution $\Sigma(p)$. As an
example we will consider the dressing of baryon resonance
$N^{\prime}$ $(J^{P}=1/2^{\pm})$  due to interaction with $\pi N$
system. Interaction lagrangian is of the form
\begin{equation}
{L}_{int}=g\overline{\Psi}^{\
\prime}(x)\gamma^5\Psi(x)\cdot\phi(x)+h.c. \quad \text{for}\quad
N^{\prime}=1/2^{+}
\end{equation}
and
\begin{equation}
{L}_{int}=g\overline{\Psi}^{\ \prime}(x)\Psi(x)\cdot\phi(x)+h.c.
\quad \text{for}\quad N^{\prime}=1/2^{-} .
\end{equation}
Isotopical indexes are irrelevant here and omitted.\\
\noindent
\underline{Positive parity baryon resonance}\\
\begin{equation}
\Sigma(p)=ig^2 \int \frac{d^4k}{(2\pi)^4}\
\gamma^5\frac{1}{\hat{p}+\hat{k}-m_N}\gamma^5
\frac{1}{k^2-m_{\pi}^2}=I\cdot A(p^2)+\hat{p}B(p^2)
\label{one-f-loop}
\end{equation}
Let us calculate the loop discontinuity through the
Landau-Cutkosky rule:
\begin{equation}
\Delta A=-\frac{ig^2m_N}{(2\pi)^2}I_{0}, \qquad \Delta
B=\frac{ig^2}{(2\pi)^2}I_{0}\frac{p^2+m_N^2-m_{\pi}^2}{2p^2}.
\label{one-f-lp}
\end{equation}
Here $I_{0}$  is the base integral
\begin{equation*}
\begin{split}
I_{0}&=\int
d^4k\delta\big(k^2-m_{\pi}^2\big)\delta\big((p+k)^2-m_N^2\big)=\theta\big(p^2-(m_N+m_{\pi})^2\big)
\frac{\pi}{2}\sqrt{\frac{\lambda\big(p^2,m_N^2,m_{\pi}^2\big)}{\big(p^2\big)^2}},
\end{split}
\label{ibase}
\end{equation*}
and $\lambda\big(a,b,c\big)=\big(a-b-c\big)^2-4bc$.

Parity conservation tells us that in the transition
$N^{\prime}(1/2^+)\to N(1/2^+)+\pi(0^{-})$ the $\pi N$ pair has
the orbital momentum $l=1$. But according to threshold
quantum-mechanical theorems, the imaginary part of a loop should
behave as $q^{2l+1}$ at $q\to0$, where $q$ is momentum of $\pi N$
pair in CMS. However, we see that this property does not hold for
$A, B$ components. But calculating the imaginary part of
$\bar{\Sigma}^{M}$ components
\begin{equation}
Im\ \bar{\Sigma}^{1}= Im\ \big(A+\sqrt{p^2}B\big)\sim q^3, \quad
\quad Im\ \bar{\Sigma}^{2}= Im\ \big(A-\sqrt{p^2}B\big)\sim
q^1,
\label{im-ofloop}
\end{equation}
we can see that the $\bar{\Sigma}^{1}$ demonstrates
the proper threshold behavior.\\
\noindent
\underline{Negative parity baryon resonance}\\
\begin{equation}
\begin{split}
\bar{\Sigma}(p)&=ig^2 \int \frac{d^4k}{(2\pi)^4}\
\frac{1}{\hat{k}+\hat{p}-m_N}\cdot\frac{1}{k^2-m_{\pi}^2}=IA(p^2)+
\hat{p}B(p^2),\\
\Delta A&=-i\frac{g^2m_N}{(2\pi)^2}I_{0},\qquad \Delta
B=\frac{-ig^2}{(2\pi)^2}I_{0}\frac{p^2+m_N^2-m_{\pi}^2}{2p^2}
\end{split}
\label{rec-of-loop}
\end{equation}
Imaginary parts of $\bar{\Sigma}^{1,2}$ again exhibit correct
threshold behavior
\begin{equation*}
 Im\ \bar{\Sigma}^{1} \sim q^1, \quad \quad
 Im\ \bar{\Sigma}^{2}  \sim q^3 .
\end{equation*}

The considered examples show  that only $\bar{\Sigma}^{1}$
component, which has the pole \mbox{$1/(\sqrt{p^2}-m)$}
demonstrates the proper parity. Another component
$\bar{\Sigma}^{2}$, which has the pole \mbox{$1/(-\sqrt{p^2}-m)$},
demonstrates the opposite parity (antifermion!).

\noindent {\bf 3.2}\ \
 Let us consider the nearest analogy to the
R.--S. field: the joint dressing of two fermions of opposite
parity $1/2^{\pm}$. We will suppose that interaction conserves the
parity. Now the Dyson-Schwinger equation has the matrix form
\begin{equation}
G_{ij}=\big(G_{0}\big)_{ij}+G_{ik}\Sigma_{kl}\big(G_{0}\big)_{lj},
\qquad i,j,k,l=1,2. \label{two-f-DS}
\end{equation}
Every element in this equation has $\gamma$-matrix indexes which
are omitted.

Decomposition of propagator now is of the form (compare with
(\ref{p2-expan}))
\begin{equation}
S(p)=\sum_{M=1}^{4}\mathcal{P}_{M}\bar{S}^{M}, \ \ \ \
\mathcal{P}_{1}=\Lambda^{+},\quad
\mathcal{P}_{2}=\Lambda^{-},\quad
\mathcal{P}_{3}=\Lambda^{+}\gamma^5,\quad
\mathcal{P}_{4}=\Lambda^{-}\gamma^5. \label{p4-expan}
\end{equation}

Multiplicative properties of this basis are seen from
Table~\ref{mult1}. The Dyson-Schwinger equation \eqref{two-f-DS}
reduces for equation on the coefficients $\bar{G}^{M}$:
\begin{equation}
\Big(\sum_{M=1}^{4}\mathcal{P}_{M}\bar{G}^{M}\Big)
\Big(\sum_{L=1}^{4}\mathcal{P}_{L}\bar{S}^{L}\Big)=
\mathcal{P}_{1}+\mathcal{P}_{2},
\end{equation}
where  $\bar{G}_{M}$, $\bar{S}_{L}$ are the matrices $2\times 2$.
It leads to matrix equations:
\begin{eqnarray}
\begin{split}
&G_{1}S_{1}+G_{3}S_{4}=E_{2},\hspace{10mm}  G_{2}S_{2}+G_{4}S_{3}=E_{2},\\
&G_{1}S_{3}+G_{3}S_{2}=0,\hspace{12mm}  G_{4}S_{1}+G_{2}S_{4}=0,
\end{split}
\end{eqnarray}
where $E_{2}$ is the unit matrix  $2\times 2$. Solutions:
\begin{equation}
\label{p4-DS-split-sol}
\begin{split}
G_{1}&=\Big[S_{1}-S_{3}\big(S_{2}\big)^{-1}S_{4}\Big]^{-1},
\hspace{22mm}
G_{2}=\Big[S_{2}-S_{4}\big(S_{1}\big)^{-1}S_{3}\Big]^{-1},\\
G_{3}&=-\Big[S_{1}-S_{3}\big(S_{2}\big)^{-1}S_{4}\Big]^{-1}S_{3}\big(S_{2}\big)^{-1},
\ \ \ \
G_{4}=-\Big[S_{2}-S_{4}\big(S_{1}\big)^{-1}S_{3}\Big]^{-1}S_{4}\big(S_{1}\big)^{-1}.
\end{split}
\end{equation}

\noindent {\bf 4.}\ \ Comparing Tables \ref{tt} and \ref{mult1},
one can conclude that presence of the nilpotent operators
$\mathcal{P}_7$ -- $\mathcal{P}_{10}$ in decomposition
(\ref{ddecomp}) is an indication for the transitions between
components of different parity $1/2^{\pm}$. To make sure in this
conclusion, we can calculate the R-S. self-energy. As an example
we will take the standard interaction lagrangian $\pi N\Delta$
\begin{equation}
L_{int}= g_{\pi N\Delta}\ \overline{\Psi}^{\ \mu}(x)
(g^{\mu\nu}+a\gamma^{\mu}\gamma^{\nu}) \Psi(x)\cdot
\partial_{\nu} \phi(x) + h.c.\ ,
\end{equation}
where a  is some arbitrary parameter.

We saw  that in case of Dirac fermions the  propagator
decomposition in basis of projection operators demonstrates the
definite parity. We can expect the similar property for R.--S.
field in $\Lambda$-basis. Calculation of self-energy contribution
\cite{KL1} leads to
\begin{eqnarray*}
\Delta \bar{J}_1= \Delta J_1 + E \Delta J_2 \sim q^3,\  & \Delta
\bar{J}_2=\Delta J_1 - E \Delta J_2 \sim q^5,        \  & \Delta
\bar{J}_3=\Delta J_3 + E \Delta J_4 \sim q^3,
\nonumber \\
 \Delta \bar{J}_4= \Delta J_3 - E \Delta J_4 \sim q, \ \  &
\Delta \bar{J}_5=\Delta J_5 + E \Delta J_6 \sim q,   \  &  \Delta
\bar{J}_6=\Delta J_5 - E \Delta J_6 \sim q^3 .
\end{eqnarray*}
Such behavior indicates that the components $\bar{J}_1,\bar{J}_2$
exhibit the spin-parity $3/2^+$, while the pairs of coefficient
$\bar{J}_3,\bar{J}_4$ and $\bar{J}_5,\bar{J}_6$ correspond to
$1/2^+$, $1/2^-$ contributions respectively.\\

\noindent {\bf 5.}\ \ Thus we obtained the simple analytical
expression \eqref{solve} for the interacting R.--S. field
propagator which accounts for all spin components. To derive it we
introduced the spin-tensor basis \eqref{L-basis} with very simple
multiplicative properties.

The obtained dressed propagator \eqref{solve} solves an algebraic
part of the problem, the following step is renormalization. Note
that the investigation of dressed propagator is the alternative
for more conventional method based on equations of motion (see,
\textit{e.g.} Ref.~\cite{Pas99} and references therein).

We found that the nearest analogy for dressing of the $s=1/2$
sector is the joint dressing of two Dirac fermions of different
parity. Some hint for such spin-parity content may be seen from
algebraical properties of $\Lambda$ basis \eqref{L-basis} with
presence on nilpotent operetors. Caculation of the self-energy
contributions in case of $\Delta$ isobar confirms it: in the
Rarita-Schwinger field besides the leading $s=3/2$ contribution
there are also two $s=1/2$ components of different parity.

This work was supported by RFBR grant 05-02-17722a.


\end{document}